\documentclass{llncs}
\pdfoutput=1
\usepackage{subfigure}
\usepackage{multirow}
\usepackage[ruled]{algorithm2e}
\usepackage{amsmath}
\usepackage{epsfig}
\usepackage{hhline}
\usepackage{tabularx}
\usepackage{comment}

\newcommand{\ordre}{$\prec$}
\newcommand\locatefig[1]{figure/dmpmccasc#1}

\begin{document}

\title{Highly Scalable Multiplication for Distributed Sparse Multivariate Polynomials on Many-core Systems}

\author{Micka\"el Gastineau\inst{1} \and Jacques Laskar\inst{1}}
\authorrunning{Micka\"el Gastineau et al.} 

\institute{IMCCE-CNRS UMR8028, Observatoire de Paris, UPMC\\
Astronomie et Syst\`emes Dynamiques\\
77 Avenue Denfert-Rochereau\\
75014 Paris, France\\
\email{gastineau@imcce.fr}, \email{laskar@imcce.fr} }

\maketitle

\begin{abstract}
We present a highly scalable algorithm for multiplying sparse multivariate polynomials represented in a distributed format. This algorithm targets not only the shared memory multicore computers, but also computers clusters or specialized hardware attached to a host computer, such as graphics processing units or many-core coprocessors. The scalability on the large number of cores is ensured by the lacks of synchronizations, locks and false-sharing during the main parallel step.

\end{abstract}

\section{Introduction}

Since the emergence of computers with multiple processors, and nowadays with several cores per processor, computer algebra systems have been trying to take advantage of such computational powers to reduce execution timings.
As sparse multivariate polynomials are intensively present in many symbolic computation problems, the algorithms of the basic operations on these objects, such as multiplication, have been designed to use the available processors in workstations. These algorithms depend on the polynomial representation in main memory. The multivariate polynomials are usually stored in a distributed or recursive format. In the distributed format, a polynomial is a list of terms, each term being a tuple of a coefficient and an exponent. In the recursive form, a polynomial is considered as an univariate polynomial whose coefficients are polynomials in the remaining variables. 

When the inputs are sparse multivariate polynomials, only the naive schoolbook product, that is the pairwise term products, is usually more optimal in practice than the asymptotically fast multiplication algorithms. Several parallel algorithms have been proposed for modern parallel hardware. An algorithm for the recursive representation has been designed using a work-stealing technique \cite{1837220}. It scales at least up to 128 cores for large polynomials. Several algorithms \cite{1576739}, \cite{biscani2012} have been designed for the distributed format for a parallel processing. The algorithm due to Monagan \cite{1576739} uses binary heaps to merge and sort produced terms but its scalability depends on the kind of the operands. Indeed, dense operands shows a super-linear scalability 
\cite{1576739}, 
 \cite{1837220}. If the number of cores becomes large, no improvement is observed because each thread should process at least a fixed number of terms of the input polynomials. This behavior is due to the fact that Monagan's algorithm benefits from the shared cache inside the processor. For sparse operands, a sub-linear speedup on a limited number of cores and a regression above these number of cores is observed 
 but their algorithm only focuses on single processor computers with multiple cores sharing a large cache. The working threads have a private heap to sort their owned results and share a global heap for computing the output polynomial. The access by a thread to the global heap requires a lock statement to avoid a race condition. This lock statement avoids to obtain a good scalability on a large number of cores. The algorithm designed by Biscani \cite{biscani2012} and implemented in the Piranha algebraic manipulator \cite{biscani2009} has no limitation according to the number of cores. The work is split in closed intervals based on the hashed value of the operands' terms and pushed in a list of available tasks. Since the result of the different tasks may overlap, the access to the two lists of available and busy tasks is controlled by a mutual-exclusion lock to avoid a race condition. This single mutual-exclusion lock becomes a bottleneck when the number of cores becomes very large. Biscani's algorithm assumes that the cost of the access to the global memory for the result is the same for all the cores and does not depend on the memory location while two threads executed on two different processors may write successively the result to the same location.

Several new hardware processing units have appeared in the last decade, such as the multi-core processors in desktop computers or laptops, GPU with hundred or thousands of elementary processing units or specialized accelerators. In these different hybrid architectures, the memory access times depend on the memory location because each processor accesses faster to its a own attached global memory. Cluster of nodes embedding all these different processors are available and may be used to perform the multiplication of sparse polynomials. 
We present a new algorithm for sparse distributed multivariate polynomial targeting these different architectures in section \ref{sec:Algorithm}. Using a small specialization of one step inside this algorithm due the constraints of the different hardwares, we adapt it to a cluster of computers in section \ref{sec:SCON} and to specialized many-core hardware in section \ref{sec:SMCH}. Benchmarks for these computers are presented in section \ref{sec:Benchmarks}.

\section{Algorithm on Shared Memory Computers}\label{sec:Algorithm}

The designed algorithm should minimize the number of synchronizations or locks between threads in order to obtain a good scalability on many cores.
 Indeed, many synchronizations or locks are required only if different threads compute the terms of the result which have the same exponent. To avoid any lock or synchronization during the computations of resulting terms, a simple strategy is that each thread computes independent terms. Computation of independent terms is very easy if a recursive data structure for the polynomials is used, as shown in Wang \cite{Wang:1996:PPO:241129.241133} and Trip \cite{Gastineau:2011:TCA:1940475.1940518} but if a distributed form is used then this task is much more tricky.
The proposed algorithm \ref{alg:mulomp} requires two major steps.
A preliminary step is required to split the work between threads to avoid any communication between the threads during the computational task. 
 
Let $c$ be the number of available cores and the same number of computational threads.
Let us consider the polynomials in $m$ variables $x_1, \dots, x_m$, $$A({\bf x})=\sum_{i=1}^{n_a}a_i{\bf x}^{\alpha_i}\text{ and }B({\bf x})=\sum_{j=1}^{n_b}b_j{\bf x}^{\beta_j}$$
where ${\bf x}$ corresponds to the variables $x_1, \dots, x_m$, the $a_i$ and $b_j$ are numerical coefficients, and the $m$-dimensional integer vectors $\alpha_i$ and $\beta_j$ are the exponents. These polynomials are stored in a sparse distributed format and their terms are sorted with a monomial order \ordre. 

The product $P$ of $A$ and $B$ is the sum of the terms $P_{i,j} = a_ib_j{\bf x}^{\gamma_{i,j}}$ where $\gamma_{i,j}={\alpha_i+\beta_j}$ for $i=1\dots n_a$ and $j=1\dots n_b$. 
 We can construct the $n_a\times n_b$ matrix of the sum of exponents, called {\it pp-matrix} following Horowitz' denomination \cite{Horowitz:1975:SAP:321906.321908}, to understand how the work is split between the threads. In fact, this matrix (Fig. \ref{ppmatrix}) is never stored in memory during the execution of the algorithm due to its size.
 
\begin{figure}
$$\begin{array}{cc}
& \begin{bmatrix}\phantom{_{,_a}}\beta_{1}\phantom{_{n}} & \dots & \phantom{_{,_a}}\beta_{j\phantom{_{n}}} & \dots & \phantom{_{,_a}}\beta_{n_b\phantom{_{n}}}\end{bmatrix} \\
& \\
\begin{bmatrix}\alpha_1 \\\dots \\\alpha_i\\\dots\\\alpha_{n_a}\end{bmatrix} &
\begin{bmatrix}\gamma_{1,1} & \dots & \gamma_{1,j} & \dots & \gamma_{1,n_b} \\
 \dots & \dots & \dots & \dots & \dots \\
 \gamma_{i,1} & \dots & \gamma_{i,j} & \dots & \gamma_{i,n_b} \\
 \dots & \dots & \dots & \dots & \dots \\
 \gamma_{n_a,1} & \dots & \gamma_{n_a,j} & \dots & \gamma_{n_a,n_b} 
 \end{bmatrix}
\end{array}$$
\caption{pp-matrix : matrix of the sum of exponents of $A\times B$}
\label{ppmatrix}
\end{figure} 

As each thread must compute independent terms,  each thread must process all the pairwise term products $P_{i,j}$ of the  {\it pp-matrix} which have the same value for $\gamma_{i,j}$. Since the possible values of all the $\gamma_{i,j}$ are in the interval $\left[\gamma_{1,1},\gamma_{n_a,n_b}\right]$, this interval may be split into subintervals which are processed by the different threads. If this interval is split in equal subintervals, the load-balancing between the threads will be very poor. Several values $\gamma_{i,j}$ of the {\it pp-matrix}  are selected and used to split into subintervals in order to obtain a better load-balancing. Since a simple and fast method to select these values does not guarantee that duplicated values are not selected,  duplicated selected values are removed and the remaining values $\gamma_{i,j}$ are the bounds of the subintervals. Left-closed, right-open subintervals are required to avoid that all the $\gamma_{i,j}$, which have the value of one of the bounds of the interval, remain in a single subinterval. The exponent $\gamma_{end}$ is introduced in order to have a same interval type (left-closed, right-open) for the last subinterval which must contain the value $\gamma_{n_a,n_b}$ and $\gamma_{end}$ is any exponent greater than $\gamma_{n_a,n_b}$ according to the monomial order. 

The algorithm begins with the construction of the set $S^\star$ which consists of the selection of $n_{s^\star}$ values (or exponents) inside $\Gamma$.  The selection method must always select the exponents $\gamma_{1,1}$ and $\gamma_{end}$ in order to be sure that all pairwise term products will be processed in the next step of the algorithm. $n_{s^\star}$ needs to be provided as an input of the algorithm and its value must remain very small in front of the size of $\Gamma$ since the first step of the algorithm needs to be fast. The way to select the $n_{s^\star}$ exponents will be discussed in section \ref{sec:CHOICE}. This set $S^\star$ is then sorted according to the monomial order and its duplicate values are removed in order to obtain the new subset $S$. The number of elements of the set $S$ is noted $n_s$. This first step is very fast and could be computed by a single thread. Of course, this step may be parallelized using a parallel sorting algorithm. If we define the set $\Gamma=\{\gamma_{i,j}\mid 1\leq i\leq n_a\text{ and } 1\leq j\leq n_b\}  {\bigcup}  \{\gamma_{end}\}$, then the first step produces the following set $$S=\{S_k \mid1\leq k\leq n_s\text{ and } S_k\in\Gamma\} \text { with }S_1=\gamma_{1,1}, S_{n_s}=\gamma_{end}\text{ and }S_k<S_{k+1}$$

After this preliminary step, every $\gamma_{i,j}$ could be located inside a single interval $\left[S_k,S_{k+1}\right[$ and all the $\gamma_{i,j}$ with the same value are located in the same interval.   The threads may process the $n_s-1$ intervals of exponents at the same time since they compute independent terms of the result. Indeed, if a thread processes the interval $\left[S_k,S_{k+1}\right[$, it computes the summation of the selected terms $P_{i,j}$ such that $S_k\leq \gamma_{i,j}<S_{k+1}$. So the second step consists in computing the resulting terms using a parallel loop over all the $n_s-1$ intervals. As each interval may have different execution times, due to a variable amount of $P_{i,j}$ involved, the work should be balanced between the cores using a number of intervals greater than the number of cores. The load-balancing may be done using a dynamic scheduling, such as work-stealing \cite{324234}, which does not require any bottleneck synchronization. 

\begin{algorithm}
\DontPrintSemicolon
\LinesNumbered
\SetKwFunction{FindEdge}{FindEdge}
\SetKwFunction{MergeSort}{MergeSort}
\SetKwFor{Forparif}{for}{${\text{ do in parallel}}$}{end}
\KwIn{$A = \sum_{i=1}^{n_a}{a_i{\bf x}^{\alpha_i}}$}
\KwIn{$B = \sum_{j=1}^{n_b}{b_j{\bf x}^{\beta_j}}$}
\KwIn{$n_{s^\star}$ : integer number of intervals}
\KwIn{monomial order \ordre}
\KwOut{$C = \sum_{k=1}^{n_c}{c_k{\bf x}^{\gamma_k}}$ }
\BlankLine
 \tcp*[l]{First step}
$S^\star \leftarrow$Compute $n_{s^\star}$ exponents $\gamma_{i,j}=\alpha_i+\beta_j$ using an almost regular grid over the {\it pp-matrix} associated to $A$ and $B$\;
$S \leftarrow$ sort $S^\star$ using the monomial order \ordre\;
remove duplicate values from $S$\;
\tcp*[l]{$S$ has now $n_s$ sorted elements}
\BlankLine
\tcp*[l]{Second step}
\BlankLine
Initialize an array $D$ of $n_s$ empty containers for the result\;
\Forparif{$k \leftarrow 1$ \KwTo $n_s-1$}{
	$(L_{min}$, $L_{max} ) \leftarrow $ \FindEdge($A,B, S_k, S_{k+1}$) \;
	$D_k \leftarrow $\MergeSort($A,B, L_{min}, L_{max}$) \;
	}
$C \leftarrow$ concatenate all containers of $D$ using ascending order\;
\caption{mul($A,B,n_{s^\star}$). Return $A\times B$ using at most $n_s^\star$ intervals. }
\label{alg:mulomp}
\end{algorithm}

During this second step, each thread needs to check if the entry $\gamma_{i,j}$ of the {\it pp-matrix} is included inside its own current interval in order to process it or not. If the thread checks each entry, each thread will perform $n_an_b$ comparisons which are very inefficient. As the {\it pp-matrix} has an ordered structure, as $\gamma_{i,j}<\gamma_{i+1,j}$ and $\gamma_{i,j}<\gamma_{i,j+1}$, this property may be exploited to find efficiently the necessary entries of the intervals. For each line of this matrix, only the location of the first and last element, which corresponds to the first and last exponent processed by the thread, should be determined. So each thread needs to find the edge of the area of terms that it should process for the current interval. This edge consists almost of two lines, which correspond to the first exponents and last exponents on each line, as shown in the figure \ref{labelfigsplitthread}. This work is done by the function FindEdge. 
This algorithm consists in storing the location of the first, respectively last, column $j$ where $S_k\leq\gamma_{i,j}$, respectively $\gamma_{i,j}<S_{k+1}$, in two arrays $L_{min}$ and $L_{max}$ of size $n_a$. Its complexity is $\mathcal{O}(n_a+n_b)$ because, when the thread processes the line $i+1$ of the matrix, it does not start at the column 1 but at the found column in the previous line $i$, as $\gamma_{i,j}<\gamma_{i+1,j}$. $n_{s^\star}$ needs to be kept small because each thread will have to process $(n_a+n_b)(n_s-1)/c$ exponents to compute these arrays if the work is well balanced. As each thread has its own arrays $L_{min}$ and $L_{max}$, the additional memory usage requirement for these arrays is only $2n_ac$ integers during the second step. However, the storage of these arrays is not required if it is possible to combine this function with the function MergeSort but this depends on the algorithm used in that function.
 
Using its own arrays $L_{min}$ and $L_{max}$, each thread computes the summation of its own terms $P_{i,j} = a_ib_j{\bf x}^{\gamma_{i,j}}$ using any sequential comparison-based sorting algorithm (function MergeSort in the algorithm) and store them in a container $D_k$ associated to the corresponding interval. No concurrent writing or reading access occurs to the same container because threads need to read or write data only about their own current interval. Johnson proposes a sequential algorithm \cite{Johnson:1974:SPA:1086837.1086847} which computes the result using a binary heap. If the multiplication produces $\mathcal{O}(n_a+n_b)$ terms, only $\mathcal{O}(n_an_b\log\min(n_a,n_b))$ comparisons of exponents are required. Monagan and Pearce have improved this algorithm with a chained heap \cite{Monagan:2010:PSP:1837210.1837227}. When all threads have finished to process all the intervals, a simple concatenation of the containers is performed to obtain the canonical form of the polynomial as the containers of $D$ are already sorted according to the sorted intervals. 

\begin{algorithm}[t]
\DontPrintSemicolon
\SetKwFunction{FindEdge}{FindEdge}
\SetKwFunction{MergeSort}{MergeSort}
\SetKwFunction{inparallel}{// do in parallel on all nodes}
\SetKwFunction{secondstep}{// Second step}
\SetKwFunction{similarsecondstep}{// similar to the step 2 of Alg. \ref{alg:mulomp}}
\SetKwFunction{addstep}{// Additional step}
\SetKwFor{Forparif}{for}{${\text{ do in parallel}}$}{end}
\KwIn{$A = \sum_{i=1}^{n_a}{a_i{\bf x}^{\alpha_i}}$}
\KwIn{$B = \sum_{j=1}^{n_b}{b_j{\bf x}^{\beta_j}}$}
\KwIn{$n_{s^\star}$ : integer number of intervals}
\KwIn{monomial order\ordre}
\KwOut{$C = \sum_{k=1}^{n_c}{c_k{\bf x}^{\gamma_k}}$ }
\BlankLine
\begin{tabularx}{\hsize}{p{2.7in}cX}
node 0 & & node 1\dots$N-1$ \\\hhline{=~=}
Perform step 1 of Alg. \ref{alg:mulomp} & & \\
&&\\
Broadcast $A$, $B$, $n_s$ and $S$ &$\Longrightarrow$ & Receive $A, B, n_s$ and $S$ \\
&&\\
\multicolumn{3}{c}{\inparallel} \\
\multicolumn{3}{l}{$(L_{min}$, $L_{max} ) \leftarrow $ \FindEdge($A,B, S_k, S_{k+1}$) for $k=1\dots n_s$} \\
\multicolumn{3}{l}{$O_k \leftarrow$ number of operations for $\left[S_k,S_{k+1}\right[$ from $(L_{min}$, $L_{max} )$} \\
&&\\
Gather $O_k$ from all nodes & $\Longleftarrow$& Send $O_k$\\
\multicolumn{2}{l}{Split $S$ in $N$ consecutive intervals $\left[S_{l_1},S_{l_2}\right[$}& \\
using the cumulative summation of $O_k$
&&\\
Send the $N$ tuples ${l_1},{l_2}$ & $\Longrightarrow$ & Receive ${l_1},{l_2}$ \\
&&\\
\multicolumn{3}{c}{\similarsecondstep} \\
\multicolumn{2}{l}{Initialize an array $D$} & Initialize an array $D_{l_1\dots l_2}$\\
\multicolumn{3}{c}{\inparallel} \\
\Forparif{$k \leftarrow l_1$ \KwTo $l_2$}{
	$(L_{min}$, $L_{max} ) \leftarrow $ \FindEdge($A,B, S_k, S_{k+1}$) \;
	$D_k \leftarrow $\MergeSort($A,B, L_{min}, L_{max}$) \;
	} \\
Gather $D_k$ from all nodes & $\Longleftarrow$& Send $D_{l_1\dots l_2}$\\
$C \leftarrow$ concatenate all containers of $D$\\
\end{tabularx}
\caption{mul($A,B,n_{s^\star}$). Return $A\times B$ using at most $n_s^\star$ intervals on a cluster of $N$ computer nodes. }
\label{alg:mulmpi}
\end{algorithm}

\section{Adaptation to Computer Cluster}\label{sec:SCON}
As the second step could be computed in independent parallel tasks, our algorithm could be easily adapted to a cluster of computational nodes. Cluster of computer nodes offers a distributed memory architecture where the access time to the memory located on the other nodes is several magnitude order greater than the access to the local memory. A message passing paradigm, such as MPI standard, should be used to perform the communications between the nodes. But a pure MPI application does not take advantage of the multiple cores available inside a node. An hybrid (multi-threading+MPI) approach must be used in order to reduce the cost of the communication and to improve the parallel scheduling of the second step of the previous algorithm. We assume that the operands are located on a single node and the result should be stored on this node. So the operands should be broadcast to the other nodes. If the operands are located on different nodes, each node broadcasts its content of the operands to the other nodes. A simple parallel scheduling could use the master-slave paradigm where a node is dedicated to be the master and other nodes request intervals to this master node, process the intervals and send the result to the master. Good load-balancing in this context requires to have many intervals which involve many communications. Furthermore, the result may generate large messages which require to use the Rendezvous protocol and imply a waiting for the other slaves.

In order to limit the number of communications, a node should process consecutive intervals and send a single result for all these sorted intervals. Inside this node, the same parallel scheduling as in the shared memory context may be chosen to distribute the work between the threads. To reduce to the minimal number of communications, the number of group of consecutive intervals is chosen to be equal to the number of nodes. But an extra step is introduced to perform a good load-balancing between all the nodes. This extra step requires to compute the number of multiplications or operations required to process each interval. The cumulative summation of the number of operations is performed in order to create the group of consecutive intervals with almost the same number of operations. The master node will also compute a part of the result. Other nodes send their results back to the master node. The algorithm \ref{alg:mulmpi} shows the processing steps required to perform the multiplication on the cluster of nodes.
 
\section{Adaptation to Specialized Many-core Hardware}\label{sec:SMCH}

GPU and other dedicated cards are able to perform general-purpose computations but they have dedicated memory. So the same adaptation as for the cluster of computers is done for the data transfer between the host memory and the GPU memory. The scheduling is easier than on the cluster since it can be done by the processor of the host computer. The values $O_k$ are not computed to perform the scheduling. If several many-core hardware are connected to the host computer, only the bounds $l_1$ and $l_2$ are sent to the different cards. As the memory transfer may be expensive, host processor may compute other data, e.g. put the result in a canonical form, in order to overlap the memory communication. 

\section{Choice of the Set $S^\star$}\label{sec:CHOICE}
The set $S^\star$ should be chosen in order to balance the work as better as possible between the intervals, even if a perfect work balancing is impossible without computing all the elements of the {\it pp-matrix}. The choice of the elements of $S^\star$ could be done using an almost regular grid over the {\it pp-matrix}. This method is very fast and very simple to implement. In order to obtain the fixed $n_{s^\star}$ elements, our grid is defined using the following rules.
$$
\begin{array}{ccl}
S^\star_k &= &\left\{
\begin{array}{ll}
\alpha_i+\beta_j &\text{for } i=1 \text{ to }n_a \text{ step }\lfloor\frac{n_a}{l}\rfloor , \\
&\text{and for } j= j_{0,i} \text{ to }n_b \text{ step }\lfloor\frac{n_b}{l}\rfloor \\
\alpha_{n_a}+\beta_j & \text{for } j=1 \text{ to }n_b \text{ step }\lfloor\frac{n_b}{l}\rfloor \\
\alpha_i+\beta_{n_b} & \text{for } i=1 \text{ to }n_a \text{ step }\lfloor\frac{n_a}{l}\rfloor \\
\end{array}
\right.\\
&& \\
\text{ with }&& 
\left\{
\begin{array}{ccl}
n_{s^\star} &=& (l+1)^2\\
j_{0,i} &=& 1+\left(\left(i/\frac{n_a}{l}\right) \mod 2\right)\lfloor\frac{n_b}{2l}\rfloor \\
\end{array}
\right. \\
\end{array}
$$

\begin{figure}
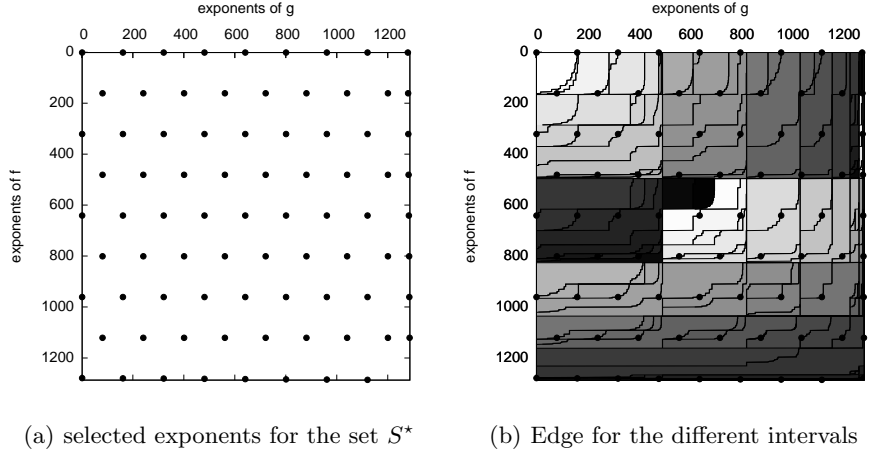

\centering
\subfigure[selected exponents for the set $S^\star$]
{
\epsfig{file=\locatefig{fig2a}, height=2.25in}
\label{labelfiggrid}
}
\subfigure[Edge for the different intervals]
{
\epsfig{file= \locatefig{fig2b}, height=2.25in}
\label{labelfigsplitthread}
}
\caption{Grid and intervals computed for the multiplication of $f=(1+x+y+2z^2+3t^3+5u^5)^{8}$ by $g=(1+u+t+2z^2+3y^3+5x^5)^{8}$ for $l=8$ or $n_{s^\star}=81$.}
\end{figure}

The value $\lfloor\frac{n_a}{l}\rfloor$and $\lfloor\frac{n_b}{l}\rfloor$ are the distances between two selected points on the same line or column in the {\it pp-mtarix}. The value $j_{0,i}$ is used to avoid the effect of the large strip. Due to the integer division, some elements in the last column and in the last line are selected to avoid a too large strip in the last part of the matrix. Figure \ref{labelfiggrid} shows an example of a computed grid and the figure \ref{labelfigsplitthread} shows the edge of the intervals computed by the different threads. Other sort of grids may be used instead of our selected grid but they have insignificant impact on the performance of the algorithm. For  example, the {\it pp-matrix} may be divided in $n_{s^\star}$ submatrices and a random exponent may be chosen inside each submatrix. Instead of selecting equidistant points on the line $i$ of our grid,  non-equidistant points may be selected on the line $i$ to generate another sort of grid. We have tested these grids and the differences of the execution time of the product are less than 1.5\% on a multicore multiprocessor computer using these grids.
 
To achieve maximal performance, the value $n_{s^\star}$ or $l$ should be chosen dynamically according to the number of available cores and/or to the number of terms of the polynomials. As $n_{s^\star}$ should remain small in order to reduce the time spent in the first step and in the function FindEdge of the algorithm, the parameter should be fitted only to the number of available cores. The parameter $l$ is preferred instead of $n_{s^\star}$ for the tuning because a simple linear variation on this parameter is possible. Its value must be tuned only once, for example at the installation of the software. Of course, for small polynomials, the tuned value $l$ may be too large and must be reduced in order to have enough work for each thread.

\section{Benchmarks}\label{sec:Benchmarks}

Three examples are selected to test the implementation of our algorithm. The two first examples are due to Fateman in \cite{844080} and Monagan and Pearce in \cite{1576739}. 

\begin{itemize}
\item Example 1 : $f_1\times g_1$ with $f_1=(1+x+y+z+t)^{40}$ and $g_1=f_1+1$. This example is very dense. $f_1$ and $g_1$ have 135751 terms and the result contains 1929501 terms.
\item Example 2 : $f_2\times g_2$ with $f=(1+x+y+2z^2+3t^3+5u^5)^{25}$ and $g_2=(1+u+t+2z^2+3y^3+5x^5)^{25}$. As shown in \cite{1576739} and \cite{1837220}, a linear speedup is quite difficult to obtain on this very sparse example.  $f_2$ and $g_2$ have 142506 terms and the result contains 312855140 terms.
\item Example 3 : $f_3\times g_3$ with $f_3=(1+u^2+v+w^2+x-y^2)^{28}$ and $g_3=(1+u+v^2+w+x^2+y^3)^{28}+1$. $f_3$ and $g_3$ have 237336 terms and the result contains 144049555 terms.
\end{itemize}

The scalability of our algorithm depends on the number of intervals $n_{s^\star}$, the size of operands ($n_a, n_b$), and the number of cores ($c$) available on the computer.
We have implemented two kinds of MergeSort algorithms for the parallel step to show its independence with respect to this algorithm. As in Monagan and Pearce, a chained heap algorithm is implemented to perform the summation of the terms but it does not include any lock as the binary heap is accessed only by one thread. This algorithm is noted {\it heap} in the tables and figures. The second sorter algorithm, noted {\it tree}, uses a tree data structure in which each internal node has exactly 16 children. At each level of this tree, four bits of the exponents are used to index the next children. If the exponents are encoded on $2^d$ bits, our tree will have $2^{d-2}$ levels. The tree associated to each interval is converted to a distributed representation at the end of the algorithm in order to obtain a canonical distributed form of the polynomial. This container uses more memory but its complexity to insert all the elements is only in $\mathcal{O}(2^{d-2}n_an_b)$. This practical complexity is better if the exponents are packed since many inserted terms have common bits inside their exponents. The exponents of the polynomials are packed on a 64-bit unsigned integer in the implemented algorithm.

To fit the value $l$, and so $n_{s^\star}$, on the available hardware, we generate randomly two sets of 280 sparse polynomials in several variables with different numbers of terms. The number of variables of these polynomials is from 4 to 8 and the number of terms varies from 10000 to 60000 terms. The products of the two sets of polynomials are performed with different values of $l$. An histogram is built with the values of $l$, whose time of execution does not differ more than 10\% from the best time for each product. 

\subsection{Shared Memory Multiprocessors}
As processors with multiple cores are now widely available in any computer, the three examples are executed on a computer with 8 Intel Xeon processors X7560 running at 2.27Ghz under the Linux operating system. Each processor has 8 physical cores sharing 24 Mbytes of cache. This computer has a total of 256Gbytes of RAM shared by its 64 cores. The parallel dynamic scheduling of the second step of the algorithm is performed by the OpenMP API \cite{openmp08} of the compiler. As the memory management could be a bottleneck in a multi-threading multiplication of sparse polynomials, the memory management is performed by the Intel threading building blocks library \cite{Reinders:2007:ITB:1461409}, noted {\it TBB}, which provides a scalable allocator instead of the operating system C library. 

\begin{figure}
\centering
\epsfig{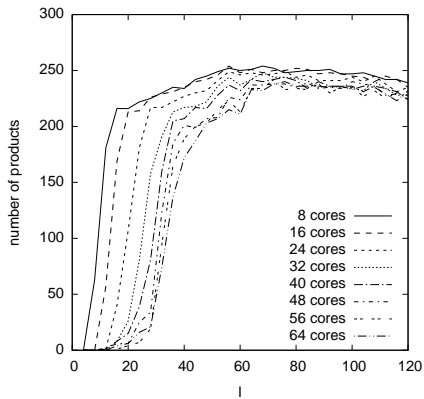}
\caption{Number of products of the two sets of 280 randomly generated sparse multivariate polynomials using different values $l$, whose execution time does not differ more than 10\% from the best time.}
\label{tuneomp}
\end{figure}

The first step has been to tune the parameter $n_{s^\star}$ or $l$ on this hardware. Figure \ref{tuneomp} shows the histogram of the number of best execution time according to the parameter $l$ using the {\it heap }algorithm. For small value of $l$, not enough parallelism is provided to get good execution time. We fix the parameter $l$ to 64 in order to perform the benchmarks on this computer. 
 
Our algorithm, noted {\it DMPMC}, is compared to the computer algebra systems Maple 16 \cite{Monagan:2011:SPM:1940475.1940521}, Piranha \cite{biscani2009} and Trip \cite{Gastineau:2011:TCA:1940475.1940518}. In all these software excepted Piranha, the coefficients of the polynomials are represented with integers using a mixed representation. For the integers smaller than $2^{63}-1$ on 64-bit computers, hardware integers are used instead of integers' type of the GMP library \cite{GMP}. The multiplication and additions of the terms use a three word-sized integers accumulator (a total of 192 bits) for the small integers. The same optimization is used in Maple \cite{Monagan:2010:PSP:1837210.1837227} and Trip. The timings for Maple 16 are the timings reported by the multiplication step of the SDMP which excludes the DAG reconstruction of the polynomial.
Piranha uses only the GMP integers and allocates memory with the same scalable memory allocator {\it TBB}. Two times are reported for Trip. The {\it dense} time is for the optimized dense recursive polynomial data structure (POLPV) and the {\it sparse} time is for the optimized sparse recursive polynomial data structure (POLYV). 

\setlength{\tabcolsep}{4pt}
\begin{table}
\caption{Execution time in seconds of the examples on the shared memory computer with integer coefficients. DMPMC uses the tuned parameter $l=64$ or  $n_{s^\star}=3636$.}
\begin{center}
\begin{tabular}{|lc|r|r|r|r|r|r|r|r|r|}
\hline
 \multicolumn{2}{|c|}{Software} & \multicolumn{3}{c|}{example 1} & \multicolumn{3}{c|}{example 2} & \multicolumn{3}{c|}{example 3} \\
\hline
&& \multicolumn{3}{c|}{cores} & \multicolumn{3}{c|}{cores} & \multicolumn{3}{c|}{cores} \\ 
&& \multicolumn{1}{c}{1}& \multicolumn{1}{c}{16}& \multicolumn{1}{c|}{64} & \multicolumn{1}{c}{1}& \multicolumn{1}{c}{16}& \multicolumn{1}{c|}{64} & \multicolumn{1}{c}{1}& \multicolumn{1}{c}{16}& \multicolumn{1}{c|}{64} \\ \hline
\multirow{2}{*}{DMPMC} &heap & 1843.2 & 116.2 & 32.3 &1317 & 83.8 & 23.9 & 2081 & 126.1 & 35.0\\ 
 & tree & 878.5 & 55.4 & 16.1 & 1394 & 90.4 & 30.2 &1632 & 102.6 & 29.7 \\ \hline
Maple 16 & & 1226.7 & 358.5 & 262.4 & 1364 & 625.0 & 900.5 & 3070 & 317.4 & 609.9\\ \hline
Piranha&  & 677.5 & 57.0 & 45.9 & 1576 & 138.4 & 174.6 & 2466 & 826.8 & 816.4 \\ \hline
\multirow{2}{*}{Trip 1.2} & dense & 649.9 & 40.3 & 10.4 &1227 & 75.6 & 19.8 & 2738 & 164.9 & 42.9 \\
 & sparse & 705.5 & 43.7 & 11.5 &1071 & 65.6 & 19.9 &2874 & 177.7 & 45.8 \\ \hline
 \end{tabular}
\end{center}
\label{tablemulopenmpdense}
\end{table}

\begin{figure}
\centering
\epsfig{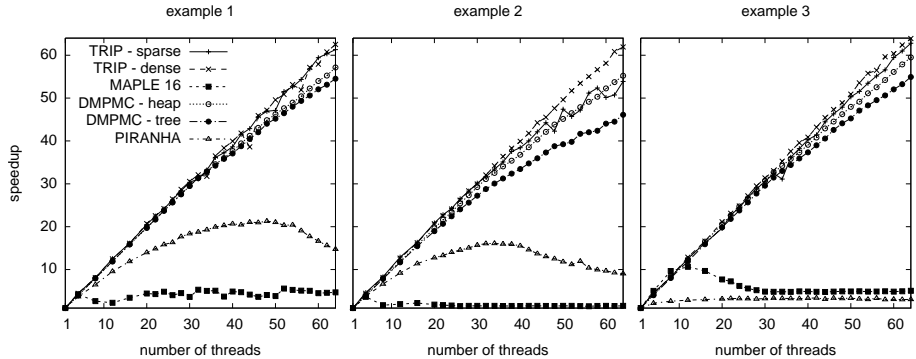}
\caption{Speedup on the shared memory computer with integer coefficients' polynomial.}
\label{exomp}
\end{figure}

Table \ref{tablemulopenmpdense} shows the execution times of the three examples on the 64 cores computer. Even if our chained heap is less tuned for the dense polynomials on single core, our algorithm for distributed representation scales with the same behavior as the recursive algorithms of Trip. We define the speedup as $T_1/T_c$ and the efficiency as $T_1/(c\times T_c)$ where $T_c$ is the execution time on $c$ cores. Figure \ref{exomp} shows the speedup for the different implementations and confirms that the SDMP algorithm of Maple 16 \cite{Monagan:2011:SPM:1940475.1940521} focuses only on a single multi-core processor with large shared memory cache. Indeed, the efficiency of the SDMP algorithm drops to less than 0.6 above 16 cores for the three examples while the efficiency of our algorithm remains above 0.7 on 64 cores. The limited scalability of Piranha is confirmed due to the access to its two global lists shared between all the threads. The kind of MergeSort algorithm has only a significant impact on the execution time but not on the scalability of our algorithm. The small differences observed on the speedup between the {\it tree}Ê and {\it heap}Ê algorithm come from the different number of memory allocations since the {\it tree}  version requires more memory allocation. Although we have used a sequential implementation of the first step, its duration remains insignificant. The computation of the edge (function FindEdge) by each thread takes only a few percent of the total time.

\subsection{Distributed Memory Computers}

\begin{figure}
\centering
\label{exmpi}
\epsfig{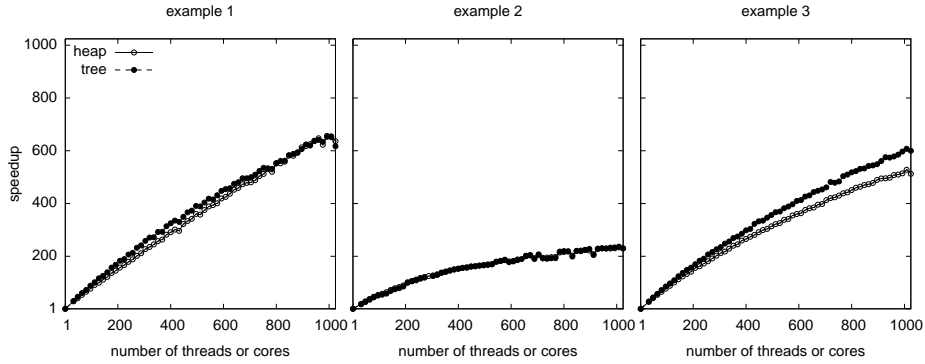}
\caption{Speedup of the DMPMC algorithm on a cluster of 64 nodes (a total of 1024 cores) interconnected with a QDR InfiniBand network. The coefficients of polynomials are 
quadruple precision floating point numbers.}
\end{figure}

In this second set of experiments, algorithm \ref{alg:mulmpi} is implemented using the hybrid approach OpenMP+MPI on the MesoPSL cluster with 64 nodes interconnected with a QDR InfiniBand network for a total of 1024 cores. Each node embeds two Intel E5-2670 processors sharing a total of 64 Gbytes of RAM between the 16 cores of the node. Quadruple precision floating point numbers have been used for the polynomials coefficients instead of variable size integers coefficients in order to simplify the exchange of the coefficients between the nodes. Figure \ref{exmpi} shows the speedup of the algorithm on this cluster with the tuned parameter $n_{s^\star}=8\sqrt{c}$ where $c$ is the total number of cores. The speedup is defined as $T_{1,\text{OpenMP}}/T_{n,\text{MPI}}$ where $T_{1,\text{OpenMP}}$ is the execution time on one core of a single node using the OpenMP implementation of the algorithm \ref{alg:mulomp} and $T_{n,\text{MPI}}$ is the execution time on $n$ nodes using the hybrid OpenMP+MPI implementation of algorithm \ref{alg:mulmpi}. The limitation of the 2 GB maximum message size in the MPI implementation of the cluster requires to implement a custom gather operation using the send/receive messages in order to collect the result on the root node which have significant impact on the timings. This bottleneck is especially visible on the large result of the second example which contains more than 300 million terms since half of the time is spent to transfer the result from the slave nodes to the master node. However, the algorithm always continues to scale according to the number of available cores. In all cases, the algorithm scales well up to at least two hundred cores.  Similar behaviors are obtained if double precision coefficients are used instead of quadruple precision coefficients.

\setlength{\tabcolsep}{3pt}
\begin{table}
\caption{Execution time in seconds and speedup of the examples on the Nvidia Tesla S2050 with double precision floating-point coefficients. Each example is executed with a different number of threads in a group with a tuned value of the parameter $l$ for the corresponding group. }
\begin{center}
\begin{tabular}{|c|c|c|c|c|c|c|c|c|c|c|c|c|c|}
\hline
group of & $l$&\multicolumn{4}{c|}{example 1} & \multicolumn{4}{c|}{example 2} & \multicolumn{4}{c|}{example 3} \\
$t$ threads & & \multicolumn{4}{c|}{} & \multicolumn{4}{c|}{} & \multicolumn{4}{c|}{} \\ \hline
&& \multicolumn{4}{c|}{gpu} & \multicolumn{4}{c|}{gpu} & \multicolumn{4}{c|}{gpu} \\ 
&& \multicolumn{1}{c}{1}& \multicolumn{1}{c}{2}& \multicolumn{1}{c}{3} &\multicolumn{1}{c|}{4}& \multicolumn{1}{c}{1}& \multicolumn{1}{c}{2}& \multicolumn{1}{c}{3} &\multicolumn{1}{c|}{4} & \multicolumn{1}{c}{1}& \multicolumn{1}{c}{2}& \multicolumn{1}{c}{3} &\multicolumn{1}{c|}{4} \\ \hline
\multirow{2}{*}{32}& \multirow{2}{*}{90}& 42 & 24 & 19 & 16 & 98 & 61 & 43 & 36 & 181 & 103 & 83 & 67 \\ 
& & \hphantom{1.0x} &1.7x&2.2x&2.5x& \hphantom{1.0x} &1.6x&2.2x&2.7x& \hphantom{1.0x} &1.8x&2.2x&2.7x\\ \hline
\multirow{2}{*}{64} & \multirow{2}{*}{90} & 32 & 18 & 14 & 12 & 71 & 42 & 30 & 25 & 129 & 71 & 55 & 44 \\ 
&& \hphantom{1.0x} &1.7x&2.3x&2.7x& \hphantom{1.0x} &1.7x&2.3x&2.8x& \hphantom{1.0x} &1.8x&2.3x&2.9x\\ \hline
\multirow{2}{*}{128}& \multirow{2}{*}{60} & 35 & 20 & 15 & 13 & 73 & 42 & 31 & 25 & 141 & 79 & 63 & 49 \\ 
&& \hphantom{1.0x} &1.7x&2.3x&2.7x& \hphantom{1.0x} &1.7x&2.4x&2.8x& \hphantom{1.0x} &1.8x&2.2x&2.9x\\ \hline
\multirow{2}{*}{256}&\multirow{2}{*}{40} & 40 & 22 & 17 & 14 & 79 & 43 & 32 & 26 & 149 & 91 & 68 & 57 \\ 
&& \hphantom{1.0x} &1.8x&2.3x&2.7x& \hphantom{1.0x} &1.8x&2.5x&  3x& \hphantom{1.0x} &1.6x&2.2x&2.6x\\ \hline
\multirow{2}{*}{512}&\multirow{2}{*}{25} & 48 & 29 & 23 & 20 & 93 & 52 & 38 & 31 & 174 & 106 & 77 & 67 \\ 
&& \hphantom{1.0x} &1.6x&  2x&2.4x& \hphantom{1.0x} &1.8x&2.5x&2.9x& \hphantom{1.0x} &1.6x&2.3x&2.6x\\ \hline
 \end{tabular}
\end{center}
\label{exgpu}
\end{table}
\setlength{\tabcolsep}{4pt}

\begin{table}
\caption{Execution time in seconds of the examples on the shared memory computer with double precision floating-point coefficients. DMPMC uses the tuned parameter $l=64$ or  $n_{s^\star}=3636$.}
\begin{center}
\begin{tabular}{|lc|r|r|r|r|r|r|r|r|r|}
\hline
 \multicolumn{2}{|c|}{Software} & \multicolumn{3}{c|}{example 1} & \multicolumn{3}{c|}{example 2} & \multicolumn{3}{c|}{example 3} \\ \hline
&& \multicolumn{3}{c|}{cores} & \multicolumn{3}{c|}{cores} & \multicolumn{3}{c|}{cores} \\ 
&& \multicolumn{1}{c}{8}& \multicolumn{1}{c}{16}& \multicolumn{1}{c|}{32} &\multicolumn{1}{c}{8}& \multicolumn{1}{c}{16}& \multicolumn{1}{c|}{32} & \multicolumn{1}{c}{8}& \multicolumn{1}{c}{16}& \multicolumn{1}{c|}{32} \\ \hline
DMPMC & tree &44.7& 22.6& 11.4&69.9 & 35.6 & 18.1 & 167.5 & 84.5 & 42.6\\ \hline
Piranha& & 8.6&6.5&20.0 & 62.8& 43.2 & 59.5 & 134.4 & 128.9 & 166.9 \\ \hline
Trip 1.2 & sparse & 11.49 &5.8&3.0& 30.9 & 15.4 & 7.9 & 86.6 & 43.5 & 22.3 \\ \hline
 \end{tabular}
\end{center}
\label{excpudbl}
\end{table}

\subsection{Specialized Many-core Hardware}
To test the algorithm on a many-core hardware, the benchmarks are performed on a Nvidia Tesla S2050 computing System based on the Fermi architecture interconnected through two links to a host computer using a PCI Express 2 16x controller. The host computer is the same computer as for the shared memory benchmark. The Nvidia Tesla S2050 consists of four Fermi graphics processing units. So two GPU cards share the same links to the host controller. Each GPU embeds 14 streaming multiprocessors and 3 GBytes DDR5 memory. Each streaming processor has 32 cores running at 1.15 Ghz and is able to schedule two groups of 32 threads simultaneously. A total of 448 cores is available per GPU. The kernel function running on the GPU is implemented using the CUDA programming model \cite{Sanders:2010:CEI:1891996}.

Only the {\it tree} algorithm for the MergeSort function is implemented since the {\it heap} version is not well adapted for the CUDA programming model. Indeed, an interval is not processed by a single thread but it is processed by a group of $t$ threads, called a {\it block} in the CUDA terminology. So for each interval, a group of threads constructs a temporary tree in order to merge and sort the terms. Atomic instructions have been used in order to reduce the number of synchronization inside the group of threads. To avoid the divergence of the execution path of the threads inside a group, this one handles the terms line after line. Only $t$ elements of the array $L_{min}$ are stored in the shared memory of the GPU at the same time in order to reduce the memory consumption and to avoid access to the global memory. In our implementation, the reconstruction of the canonical distributed representation from the tree is performed on the host processor with one host thread per GPU and overlaps the computations by the GPU. A simple static scheduling is performed : $\frac{n_s}{5g}$ intervals are processed at a same time by a GPU if $g$ GPUs are used for the computations. So each GPU executes 5 times the kernel function. More sophisticated scheduling may be done by overlapping memory transfer between the host and GPU memory. As no version of the GMP library is available for the GPU side, double-precision floating point numbers have been used for the coefficients on CPU and GPU. Table \ref{exgpu} shows the execution time and speedup obtained on the GPU with different number of threads ($t$) inside the group. As the kernel function needs to be transferred on the card by the CUDA driver, the timings are the average of eight executions without two first useless executions. The group of 64 threads has the best execution time for the three examples. The scalability on several GPUs is not linear for several reasons. The major reason is that the four cards share the two links to the host. A better dynamic scheduling should be done using an optimized heterogeneous scheduler, such as StarPU \cite{CPE:CPE1631}, and the computations on the card are not overlapped by the memory transfer. For comparison, the execution time of Trip, Piranha and DMPMC on the same host computer are reported in Table \ref{excpudbl} with the same kind of numerical coefficients.

\section{Conclusions}

The proposed algorithm for the multiplication of sparse multivariate polynomials stored in a distributed format does not have any bottleneck related to the numbers of cores due to the lack of synchronization or locks during the main parallel step. But it requires a preliminary one time step to tune the size of the grid to the targeted hardware. The range of targeted processor units is wide for our algorithm. The only drawback comes from the time to transfer data between nodes on the distributed memory systems due to the limited performance of the interconnection network. 
It can use any available fastest sequential merge and sort algorithm to generate the terms of the result and can benefit from any efficient dynamic scheduling. A more appropriate algorithm for this merge and sort step may be designed for the GPU hardware in order to take into account all features of these specialized hardware.

\section*{Acknowledgements}  The authors thank the computing
centre MesoPSL of the PSL Research University for
providing the necessary computational resources for this work.

%
\bibliographystyle{splncs}
\bibliography{dmpmccasc2013} 
\end{document}